\documentclass[twocolumn,runningheads]{svjour2}
\smartqed  
\usepackage{graphicx}
\usepackage{mathptmx}   
\usepackage[authoryear]{natbib}
\journalname{Astrophysics and Space Science (CoRoT/ESTA Volume)}
\begin{document}

\title{Granada oscillation Code (GraCo)
}


\author{A. Moya 
        \and
        R. Garrido 
}


\institute{Instituto de Astrof\'{\i}sica de Andaluc\'{\i}a- CSIC \at
              Cno. Bajo de Huetor, 50, Granada, Spain\\
              Tel.: +34 958121311\\
              \email{moya@iaa.es}           
}

\date{Received: date / Accepted: date}

\maketitle
\begin{abstract}
Granada oscillation code (GraCo) is a software constructed to compute
adiabatic and non-adiabatic oscillation eigenfunctions and
eigenvalues. The adiabatic version gives the standard numerical
resolution, and also the Richardson extrapolation, different sets of
eigenfunctions, different outer mechanical boundary conditions or
different integration variables. The non-adiabatic version can include
the at\-mosphere-pulsation interaction. The code has been used for
intensive studies of $\delta$ Scuti, $\gamma$ Doradus, $\beta$ Ceph.,
SdO and, SdB stars. The non adiabatic observables ``phase-lag'' (the
phase between the effective temperature variations and the radial
displacement) and ${\delta T_{eff}\over T_{eff}}$ (relative surface
temperature variation) can help to the modal identification. These
quantities together with the energy balance (``growth rate'') provide
useful additional information to the adiabatic resolution
(eigenfrequencies and eigenfunctions).  
\end{abstract}

\section{Introduction}\label{intro}

GraCo \citep{moya} is a software developed to solve non-radial
adiabatic and non-adiabatic oscillation equations. It is written in
fortran95 language. It can be used for models all over the HR
diagram. GraCo is able to work with three different sources of
equilibrium models: CESAM \citep{morel}, Granada Code \citep{claret}
and JMSTAR \citep{mac}. The numerical technique used is the so called
Henyey relaxation method as it is described in \citet{unno}
(Section 18.2). The simple representation of the system of differential
equations in terms of second-order centred differences is adopted for
the numerical resolution.

\section{Adiabatic case}\label{adia}

The adiabatic system of differential equations is described in
\citet{unno} (p. 161). The code has the possibility of choosing
between two sets of eigenfunctions. Both sets include the radial
displacement ($\xi_r$) and the Eulerian perturbation of the
gravitational potential (${1\over g}\phi^\prime$). The sets differ in
the use of the Lagrangian or the Eulerian variation of the pressure,
and the addition or not of a function of the radial displacement
($U\xi_r / r$, with $U=d\ln m / d \ln r$, $m$ the mass and $r$ the
radius), to the derivative of the Eulerian perturbation of the
gravitational potential ($d\phi^\prime / dr$) \citep{vor}.

As boundary conditions GraCo uses those prescribed in \citet{unno}
(pp 162 {\em ff}). The solutions must satisfy regularity conditions at
the innermost mesh-point. As a first surface condition the continuity of
$\phi^\prime$ and its first derivative are imposed. As second surface
condition, the mechanical one, the program offers two possibilities:
1) The Lagrangian variation of the pressure vanishes ($\delta p=0$),
or 2) makes use of the isothermal reflective wave boundary condition
(see \citet{unno}, pp 163 {\em ff}).

Another degree of freedom of GraCo is the variable of integration. The
program can solve the system of differential equations as a function of
the logarithm of the radius ($\ln r$) or the ratio between the radius
and the pressure (${r\over P}$). The first one largely weights the
inner regions and the second the outer ones. Depending on the
physics to be tested, the user can choose the most convenient variable.

In order to solve the eigenvalue problem, the outer boundary condition
for the gravitational potential is removed. In this case, for each
trial eigenfrequency we have a unique solution. But not all the
solutions obtained with every trial eigenfrequency fulfill the removed
boundary condition. The spectra of the system is the set of
eigenfrequencies and eigenfunctions fulfilling this outer boundary
condition.

As the system of differential equations has been replaced by a system
of centred difference equations of second order, the truncation error in
the eigenfrequencies and the eigenfunctions are of the order $N^{-2}$,
with $N$ the number of mesh points of the equilibrium model. To obtain
more accurate eigenfrequencies, the code can make use of the so called
Richardson extrapolation \citep{shiba}, a combination of the values
obtained with $N$ and $N/2$ mesh points cancelling the leading order
error.

Finally, GraCo can compute the first order rotational splitting
through the Ledoux coefficient \citep{ledoux}. The eigenfrequencies
for radial modes can be obtained in two ways: 1) Using the LAWE second
order differential equation, or 2) setting $\ell=0$ in the standard
non-radial system of equations.

\begin{figure}
\centering
  \includegraphics[width=0.48\textwidth]{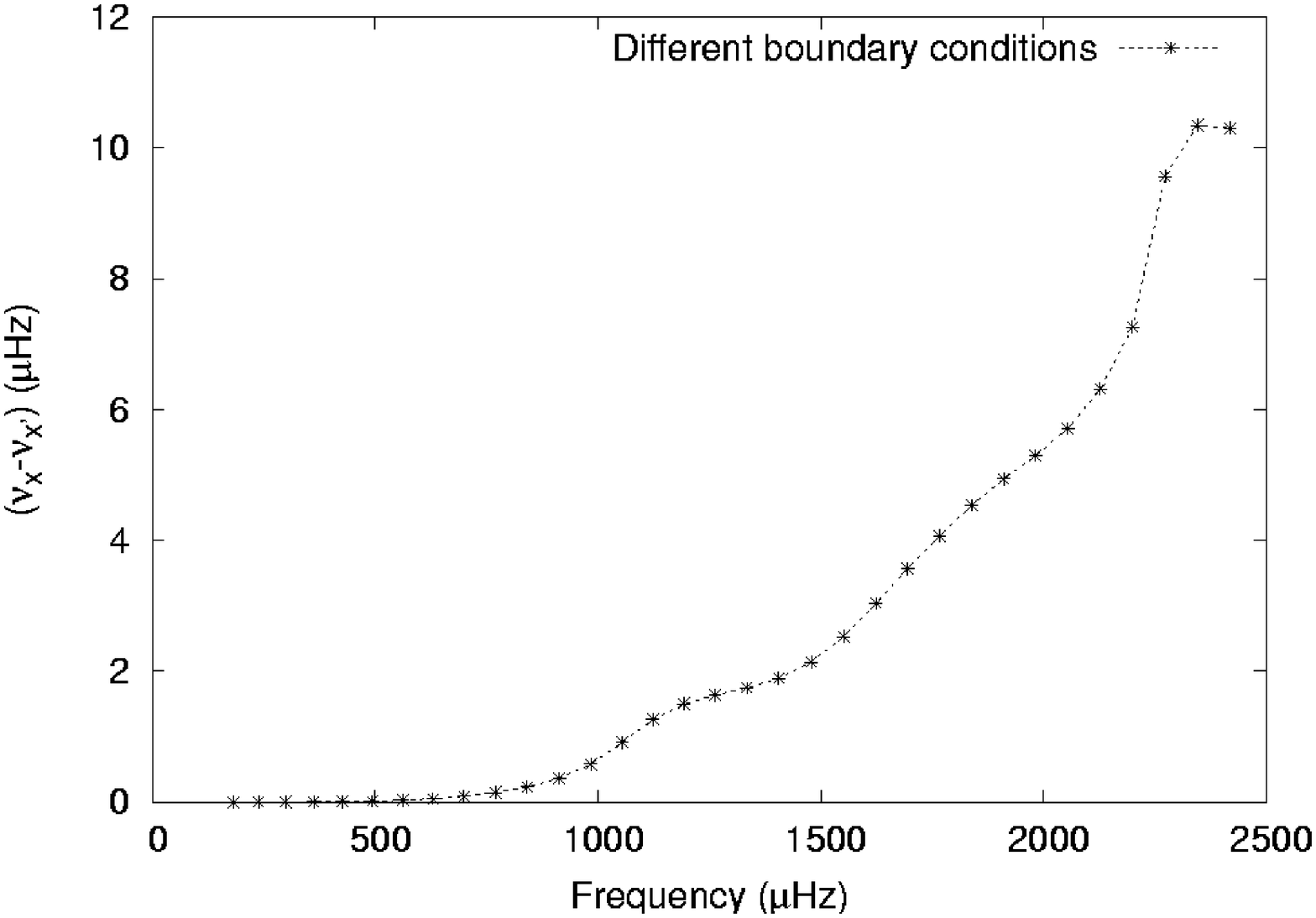}
  \rotatebox{-90}{\includegraphics[width=0.34\textwidth]{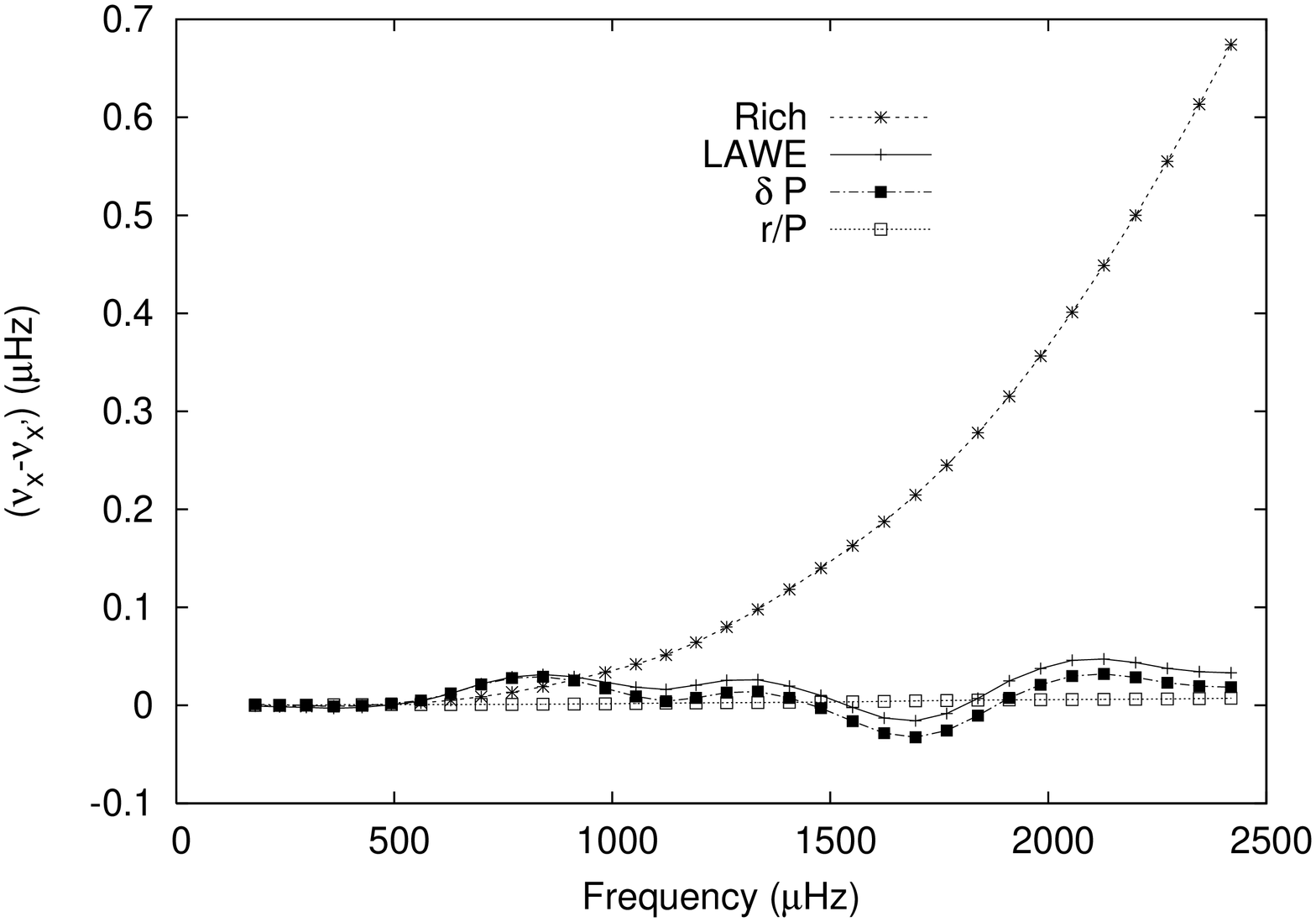}}
\caption{Frequency comparison of the different adiabatic resolutions
  given by GraCo (see text). Only radial modes are shown. The
  equilibrium model used is a $1.5M_\odot$ and $X_c=0.4$ model, with
  4042 mesh points.}
\label{1}       
\end{figure}

Fig. \ref{1} shows the adiabatic eigenfrequency differences between
all the possible aforementioned options, only radial modes are
depicted. The equilibrium model used is the last of the step1 of Task
2 \citep{task2} (this volume), with 4042 mesh points, $1.5M_\odot$
and $X_c=0.4$. As reference we have calculated eigenfrequencies with
the following options: $X=$($\ell=0$, no Richardson, $p^\prime$,
$\delta p=0$, $\ln r$). For each comparison we have changed one single
degree of freedom, remaining the rest unchanged. We show the
differences obtained in the range [200,2500] $\mu$Hz, that is, from
the fundamental radial mode to a frequency slightly larger than the
cutoff frequency (around 2250 $\mu$Hz). Top panel presents the
differences obtained when two outer mechanical boundary conditions are
used. Reference line is $\delta P=0$, and the comparison is with the
use of the isothermal reflective wave outer boundary condition. The
differences for large frequencies are of the order of units of
$\mu$Hz. These differences are similar to those obtained by other
codes (J.C. Su\'arez, private communication), but a larger study of
these differences is still needed. The use of higher order integration
procedures, as the Richardson extrapolation, do not change
significantly this differences. In bottom panel, the rest of the
comparisons are depicted. The Richardson extrapolation gives
differences of the order of tenths of $\mu$Hz, the use of $r/P$ as
integration variable provides small differences always
lower than 0.008$\mu$Hz. The LAWE differential equations and the use of
the Lagrangian variation of the pressure as eigenfunction provide
similar differences smaller than 0.05$\mu$Hz, but its profile is not
constant. For a comprehensive study of these differences see
\cite{task2} (this volume).

\section{Non-adiabatic resolution}

Additional information for asteroseismology is provided by GraCo with
the resolution of the non-adiabatic set of differential equations
described in \citet{unno} (pp 261 {\em ff}). In the non-adiabatic resolution the
eigenfrequencies and the eigenfunctions are no longer real. This makes
it possible to obtain the so called non-adiabatic observables: 1) The
``phase-lag'' ($\phi^T\equiv \phi(\xi_r)-\phi(T)$) defined as the
phase difference between the Lagrangian variation of the effective
temperature and the radial displacement. 2) The relative variation of
the Lagrangian variation of the effective temperature (${\delta
T_{eff}\over T_{eff}}$). And 3) The energy balance of each mode measured
by the ``growth rate'', directly related with the imaginary part of
the modal eigenfrequency.

The code uses the adiabatic solutions obtained for a given mode as
trial functions for the non-adiabatic relaxation procedure. The inner
boundary conditions are those described in \citet{unno} (p. 229). It
is in the outer region where GraCo presents some complexity. The code
can treat the photosphere as a boundary condition or introduce the
atmosphere-pulsation interaction resolution described by \\
\citet{dupret}. This interaction is described imposing the atmosphere
to be in thermal equilibrium and the diffusion approximation for the
radiative flux to be no longer valid. In this case two different
sets of differential equations are solved, one for the stellar core
and envelope and another for the atmosphere. A transition layer and
outer boundary conditions for the atmosphere must be defined.

\begin{figure}
\centering
  \includegraphics[width=0.48\textwidth]{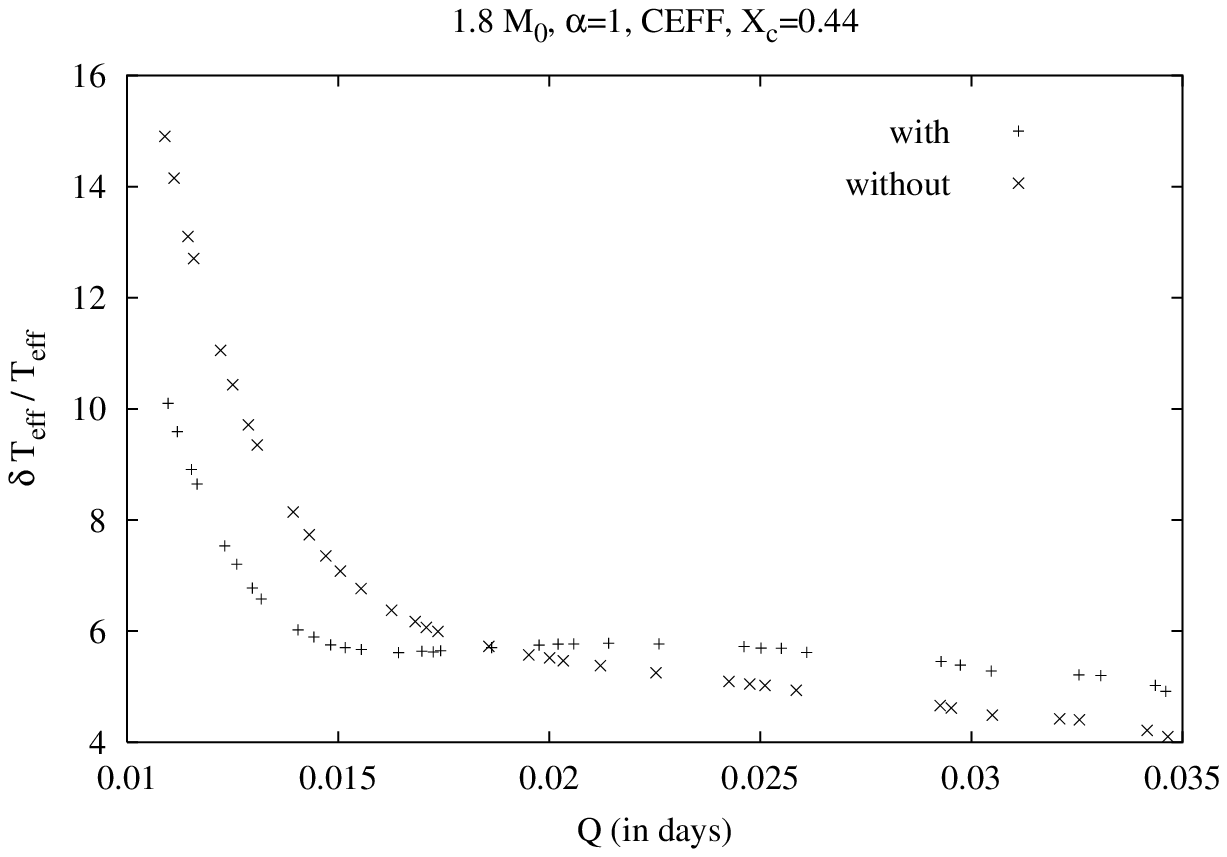}
  \includegraphics[width=0.48\textwidth]{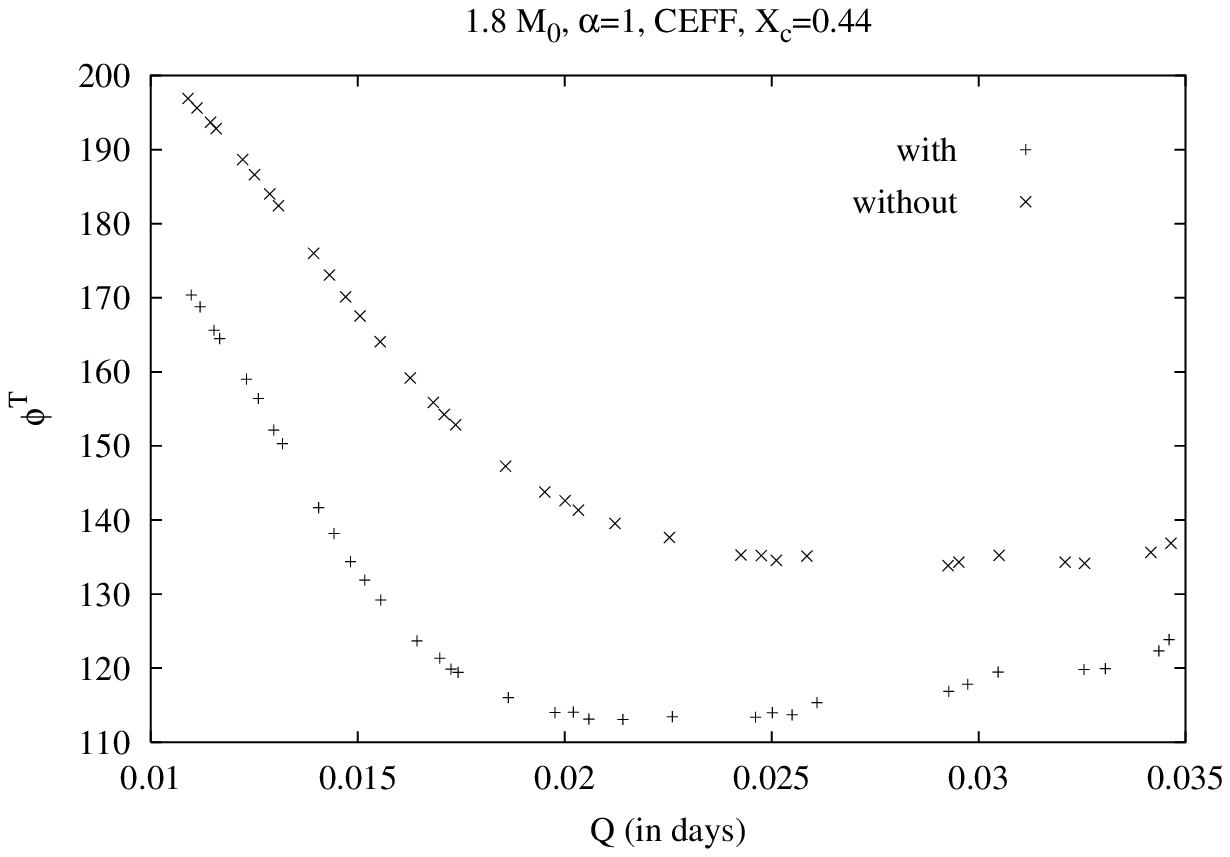}
\caption{$|\delta\mathrm{T}_{\rm eff} / \mathrm{T}_{\rm eff}|$ (top
          panel) and $\phi^T$ (in degrees, bottom panel), as a
          function of the pulsation constant $Q$ (in days) for
          different modes with spherical degrees $l=0,1,2,3$.  A model
          of a $1.8M_{\odot}$ is studied with $X_c=0.44$, a MLT
          parameter $\alpha =1$ and the CEFF equation of state.
          Results obtained ``with'' ($+$) and ``without'' atmosphere
          ($\times$) in the non-adiabatic treatment are compared.}
\label{2}       
\end{figure}

Fig. \ref{2} shows the values of the non-adiabatic observables for a
standard $\delta$ Scuti model of 1.8$M_\odot$. In this figure we can
see how the values in the case labeled ``with'' (where the
atmosphere-pulsation interaction is included) are clearly different
from those labeled ``without'' (the photosphere treated as the outer
boundary layer). This illustrates the importance of the inclusion of
the atmosphere-pulsation interaction for a better description of the
non-adiabatic observables. When the atmosphere is treated as a
boundary condition, not all the heat exchanges here are correctly
taken into account, therefore the phase-lags obtained are closer to
the adiabatic prediction ($180^\circ$). On the other hand, the
system of equations modeling the atmosphere-pulsation interaction takes into
account these non-adiabatic processes in the atmosphere, and the
resulting phase-lags are much smaller than $180^\circ$. This
atmosphere-pulsation interaction has not influence upon the growth
rate, since it takes place in layers not relevant for the driving of
the modes due to their very small density.

\begin{figure}
\centering
\rotatebox{270}{\resizebox{!}{\hsize}{\includegraphics{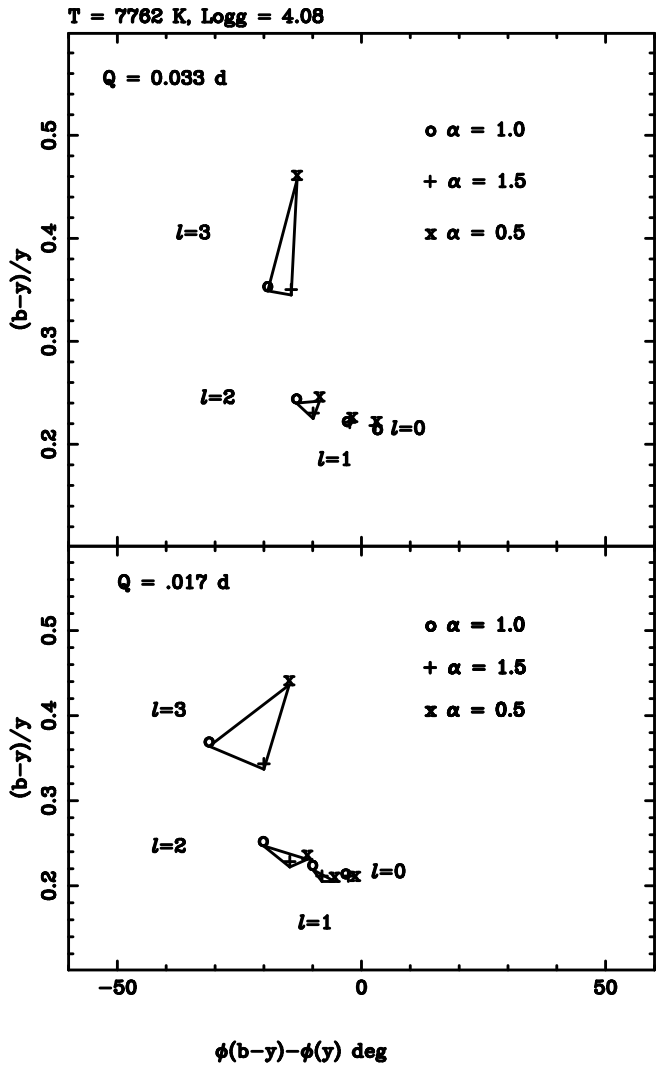}}}
\caption{The top panel shows theoretical predictions for two specific
Str\"omgren photometric bands ( ($b-y$) and $y$) for a given
theoretical model using three MLT $\alpha$ parameters in the
fundamental radial mode regime (pulsation constant near 0.033
days). The 3rd overtone regime (pulsation constant near 0.017 days) is
shown in the bottom panel. The usual observational errors for these
quantities from the ground are: $\pm 1^\circ$ for the phase
differences and $\pm 0.01$ for the amplitude ratio.}
\label{3}       
\end{figure}

All of the above-mentioned calculations have a direct influence on the
phase difference - amplitude ratio diagrams used to discriminate
oscillation modes. phase-lags, as well as relative variations in
$|\delta\mathrm{T}_{\rm eff} / \mathrm{T}_{\rm eff}|$ and $\delta
g_e/g_e$ (also calculated in the non-adiabatic resolution) can be used
to overcome the uncertainties in previous phase-ratio color
diagrams. In \citet{gar} these discrimination diagrams were made using
parametrized values for departures from adiabaticity and phase
lags. The only remaining degree of freedom is now the choice of the
MLT $\alpha$ parameter in order to describe the convection. Therefore,
discrimination diagrams depend only on this parameter, as is shown in
Fig. \ref{3} for the same equilibrium model used in fig. \ref{2}.
Theoretical predictions are plotted for two specific Str\"omgren
photometric bands (($b-y$) and $y$) using three different MLT $\alpha$
parameters in the fundamental radial mode regime (pulsation constant
near 0.033 days) and in the 3rd overtone regime (near 0.017 days).

A clear separation between the $l$-values exists for periods around
the fundamental radial. Similar behaviour is found for other modes in
the proximity of the 3rd radial overtone, although for these shorter
periods some overlapping start to appear at the lowest
$l$-values. They also show the same trend as for the fundamental radial
mode: high amplitude ratios for low MLT $\alpha$ and the spherical
harmonic $l=3$.

\section{Conclusions}

GraCo is a complete software for the resolution of systems of
differential equations related with the stellar oscillations. The
general integration scheme used in the code is the Henyey relaxation
method as it is explained in \citet{unno}. In the adiabatic frame,
different integration schemes can be used to obtain the
eigenfrequencies and eigenfunctions: 1) Two outer mechanical boundary
conditions, 2) two choices for the set of eigenfunctions, 3) two
choices of the integration variables, 4) the use or not of the
Richardson extrapolation and, 5) for radial modes the use of the LAWE
second order differential equation or set $\ell=0$ in the standard
non-radial differential equations.

On the other hand, the main characteristic of the code is that the
non-adiabatic set of differential equations can also be solved. Two
different treatments of the photosphere can be used. The first
considers the atmosphere as a single boundary layer, and the second
describes the atmosphere-pulsation interaction. This makes it possible
to obtain more accurate non-adiabatic observables (phase-lag and
${\delta T_{eff}\over T_{eff}}$) crucial for modal identification
through the multicolor photometry. On the other hand, non-adiabatic
studies also allow to study the modal energy balance, giving a
theoretical range of overstable modes.

Finally we want to remark that the analytical expressions of the
differential equations, eigenfunctions and boundary conditions, and
the numerical procedure followed in GraCo can be found in
\citet{unno}.

\begin{acknowledgements}
This work was supported by the Spanish PNE number ESP
2004-03855-C03-C01
\end{acknowledgements}


\begin{thebibliography}{3}
%
%

\bibitem[{Claret (1999)}]{claret}
Claret, A.: Studies on stellar rotation. I. The theoretical apsidal motion for evolved rotating stars. A\&A {\bf 350}, 56--62 (1999)

\bibitem[{{Dupret} {et al.} (2002)}]{dupret} {Dupret}, M.-A., {De
Ridder}, J., {Neuforge}, C., {Aerts}, C., and {Scuflaire}, R.:
Influence of non-adiabatic temperature variations on line profile
variations of slowly rotating beta Cep stars and
SPBs. I. Non-adiabatic eigenfunctions in the atmosphere of a pulsating
star. A\&A {\bf 385}, 563--571 (2002)

\bibitem[{{Garrido} {et al.} (1990)}]{gar}
{Garrido}, R., {Garcia-Lobo}, E. and {Rodriguez}, E.: Modal discrimination of pulsating stars by using Stromgren photometry. A\&A {\bf 234}, 262--268 (1990)

\bibitem[{{Ledoux}(1951)}]{ledoux}
Ledoux, P.: The Nonradial Oscillations of Gaseous Stars and the Problem
of Beta Canis Majoris. ApJ {\bf 114}, 373-- + (1951)

\bibitem[{Morel (1997)}]{morel} Morel, P.: CESAM: A code for stellar
evolution calculations. A\&Asp {\bf 124}, 597--614 (1997)

\bibitem[{{Moya} {et al.}(2003)}]{moya} Moya, A., Garrido, R., and
Dupret, M.A.: Non-adiabatic theoretical observables in {$\delta$}
Scuti stars. A\&A {\bf 414}, 1081--1090 (2004)

\bibitem[{{Moya} {et al.}(2007)}]{task2} Moya, A., et al.:
Inter-comparison of the g-, f- and p-modes calculated using different
oscillation codes for a given stellar model. This volume.

\bibitem[{{Lawlor and MacDonald}(2006)}]{mac} {Lawlor}, T.~M. and
{MacDonald}, J.: The mass of helium in white dwarf stars and the
formation and evolution of hydrogen-deficient post-AGB stars. MNRAS
{\bf 371}, 263--282 (2006)

\bibitem[{{Shibahashi} {et al.} (1981)}]{shiba}
Shibahashi, H. and Osaki, Y.: Theoretical Eigenfrequencies of Solar
Oscillations of Low Harmonic Degree L in 5-MINUTE Range. PASJ {\bf
  33}, 713-- + (1981)

\bibitem[{{Unno} {et al.} (1989)}]{unno}
{Unno}, W., {Osaki}, Y., {Ando}, H., {Saio}, H., and 
	{Shibahashi}, H.: Nonradial oscillation of stars. University
	of Tokyo Press, Tokyo (1989)

\bibitem[{{Vorontsof} {et al.} (1976)}]{vor} {Vorontsov}, S.~V. and
{Zharkov}, V.~N. and {Lubimov}, V.~M..: The free oscillations of
Jupiter and Saturn. Icarus {\bf 27}, 109--118 (1976)
\end{thebibliography}
%

\end{document}